\documentclass[aps,prb,superscriptaddress,amsmath,amssymb,preprintnumbers,twocolumn,floatfix,showpacs,showkeys,10pt]{revtex4}
\usepackage{amssymb}
\usepackage{epsfig}

\begin{document}

\title{Carrier-carrier entanglement and transport resonances  in semiconductor quantum dots}

\author{Fabrizio~Buscemi }
\email{buscemi.fabrizio@unimore.it}
\author{Paolo~Bordone}
\affiliation{CNR-INFM National Research Center on nanoStructures
and bioSystems at Surfaces ($S3$), Via Campi 213/A, 41100
Modena, Italy}
 \affiliation{ Dipartimento di Fisica, Universit\`{a} di Modena e Reggio Emilia, 41100, Modena, Italy}
\author{Andrea~Bertoni }
\affiliation{CNR-INFM National Research Center on nanoStructures
and bioSystems at Surfaces ($S3$), Via Campi 213/A, 41100
Modena, Italy}

\begin{abstract}
We study theoretically the entanglement created  in  a scattering  
between   an electron,  incoming from a source  lead, and another electron bound in the  ground state of a  quantum dot, connected to two leads. 
We analyze the role played by the different kinds of resonances  in  the transmission spectra and by the number of scattering channels,
into the amount  of quantum correlations between the two identical carriers. It is shown that the entanglement between their  energy states 
is not sensitive to the presence of Breit-Wigner resonances, while it presents a  peculiar behavior   in correspondence of Fano peaks: 
two close maxima separated by a minimum
for  a two-channel scattering, a single maximum for a multi-channel scattering.
Such  a behavior is  ascribed to the  different mechanisms  characterizing the two types  of resonances. Our results suggest that the production and detection of  
entanglement in quantum dot  structures  may be  controlled  by the manipulation of Fano resonances
through external fields.
\end{abstract}

\maketitle

\section{Introduction} 
Quantum entanglement, as one of the most spectacular features of quantum mechanics contrasting with
classical physics \cite{Eins}, has been widely investigated in the last decades, mainly because it is recognized  as a crucial
resource for quantum information processing and quantum communication \cite{Bouwn}.
It is therefore a problem of great  interest to find physical systems where the entanglement can be produced, manipulated
and detected. Recently  there have been different proposals to produce entangled states,  
such  as those based on atomic systems \cite{chan},  quantum electrodynamics  cavities \cite{Lof} and   solid state devices   \cite{bus,berbor,bor2,Oli,raa,Rasm}. 
Among the systems proposed for the realization of quantum information processing devices, those constructed with 
semiconductor  quantum dots (QDs)  are extremely promising due mainly to the controllability of their quantum state \cite{Div,Ima,Zun}.
Indeed,  semiconductor QDs posses many desirable features: an atomic-like structure, that can be fully controlled by an external electrostatical
potential; a tunable coupling to source and drain leads, which makes feasible the integration with other microelectronics
devices; and  the scalability, which seems to promise sophisticate engineering of the multi-QD structures.
 
QDs represent also an  ideal laboratory to compare exact numerical simulations of the quantum transport phenomena
with experiments, since the number of degrees of freedom involved is often small and the discreteness of QD states highly reduces
 the computational burden needed. Various effects such as
the conductance  quantization  \cite{Been}, the Coulomb blockade  due  to the electron repulsion \cite{Meira},
 the interplay between  resonances  and the charging in QD structures \cite{Wiel,Lasse}, strongly affect the transport properties. Another 
peculiar feature of the electron transmission  through QDs is  the partial retention of quantum  coherence \cite{Aik}, whose measurement,
by experimental setups exploiting quantum interference (e.g. with a QD embedded in an Aharonov-Bohm interferometer), 
may yield information about transport phenomena, not readily available from conductance measures \cite{Heib1,Heib2,Heib3}.

In the frame of   quantum   transport in semiconductors, 
theoretical and experimental investigations   have revealed   other mechanisms  governing the electronic transmission through  QD structures. In particular it has been observed
that  two kinds of resonances can be present in  conductance spectra, known as  \textit{Fano} and \textit{Breit-Wigner} resonances \cite{Gor,Fuh,Lad, Berton,Noc}.  The formers  are present
when  two transmission channels, a  resonant one and a  non-resonant one, interfere \cite{Fano}. Moreover they exhibit typical asymmetrical line shapes,
with the  transmitted phase increasing by $\pi$ on  the resonance peak and then dropping abruptly. The latters  show a symmetric line shape and may
be considered a limiting case of Fano resonances, occurring, for example, when there is  a  single  channel \cite{Breit}.

In this paper  we  address the problem of entanglement generation in a two-particle scattering  in a QD structure.
We analyze, in particular, the role played by  some mechanisms of charge  transport   in the appearance of quantum correlations.
To this aim  we consider  a  scattering event in a one-dimensional (1D) double-barrier resonant tunnelling device (that mimics the confining potential of the QD), with an 
 electron incoming  from one lead and another  electron bound in the ground state of the QD. The two particles feel the confining potential 
inside the device and interact  through  the Coulomb repulsion. Indeed carrier-carrier entanglement has  been recently investigated in various   QD structures \cite{Oli,raa,Tam,Ryc, Contre,inn, Lopez}
where  different scattering setups  are considered for the  generation of  two-electron entangled states. In the present work, we adopt a 
time-independent few-particle approach that, although computational demanding,
 can  be solved numerically to obtain the exact modulus and phase of the transmission coefficient (TC)
 of an electron crossing the charged QD. This gives us the possibility  to quantify  the  quantum correlation  between the energy states
of the scattered electron and the  bound one, and to expose its connection  with  the resonances  exhibited by the TCs  of the various 
scattering channels. Unlike previous works \cite{Oli,Lopez},  the few-particle approach we use  in this paper allows us to study the relation between the different kinds  of resonances, 
or  the number of energy levels available, with  the entanglement formation. Even if the dynamics of carriers has been  considered as 1D,
we can assume  that the results  obtained describe  a general behavior concerning also quantum transport in 2D and 3D  physical systems,
being the appearance of quantum correlations closely related  to the nature of transport resonances, whose underlying mechanisms are independent from
the dimensionality of the system.

The paper is organized  as follow. In Sec.~\ref{physicalmodel} we describe the physical model and
the numerical approach adopted  to calculate  the two-particle scattering  state and the quantum correlations  in terms of
the von Neumann entropy. In Sec.~\ref{numerical} we present   the  numerical results obtained for the entanglement in the case
of two different kinds of processes, namely two- and multi-channel scattering \footnote{Here and in the following we indicate, for brevity, as multi-channel
scattering a process with more than two channels, as detailed in Sec.~\ref{numerical}.}. Finally in Sec.~\ref{Conclusions}  we comment the results, draw final remarks
and point out issues that require further research.

\section{THE PHYSICAL MODEL AND THE NUMERICAL APPROACH}
\label{physicalmodel}
Our aim is to evaluate the entanglement between the energy states of
two electrons, interacting via the Coulomb potential, one bound in a QD and the other passing through it.
Here we  summarize the  physical model adopted for
the open QD and the numerical approach used to evaluate
the two-particle scattering state and the entanglement.

\begin{figure}[h]
  \begin{center} 
    \includegraphics[width=0.8\linewidth]{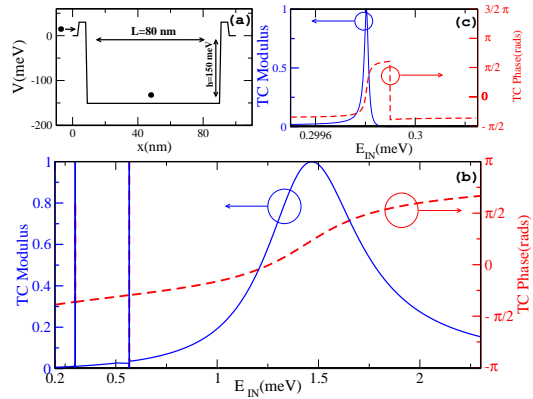}
   \caption{\label{fig:1} (Color online) Bottom panel (b):  TC as  a function of the initial energy of the incoming electron $E_{IN}$. The solid line represents the modulus of the TC and refers  to the left ordinate axis;
the dashed line represents the phase of TC and refers  to the right axis. Top panels: (a)   profile of the potential $V(x)$ in the  scattering region: the  potential well
is 150 meV deep and 80 nm wide and is connected to the leads through two 10 nm barriers of 30 meV; (c)    zoom   of  the TC  modulus and phase   for $E_{IN}\simeq  0.3$ meV where a Fano resonance occurs.}
 \end{center}
\end{figure}

We consider a quasi 1D double-barrier resonant tunnelling device as, for example, the ones formed by material
modulation in a ultrathin cylindrical nanowire \cite{Bjork}.  The
transversal dimensions  of the structure are  small compared to the
other lenghtscales so that a single transversal subband is accessible
to the carriers and the effective dynamics can be considered 1D \cite{Fogler}.  Two
small potential barriers separate the QD region from the two contacts,
as depicted in   Fig.~\ref{fig:1}(a).
The $N$ bound states and energies of the QD will be indicated as
$\chi_n$ and $E_n$ (with $n=0,1,\dots,N$ in order of increasing
energy), respectively.  

A single electron is in the QD ground state $\chi_0$,  whereas a second
electron is incoming from the left lead with  energy $E_{IN}$,  and it is
scattered by the structure potential $V_s$ and by the Coulomb
interaction with the bound particle.  The potential $V_s$  is supposed to be constant outside the
region of interest of length $\Delta$ and, without loss of generality, it
is taken to be zero in both left and right contacts.  We will consider
only cases in which the energy of the incoming electron is not
sufficient to ionize the QD, i.e. $(E_{IN}+E_0)<0$. This means that when
an electron leaves the scattering region, either reflected or
transmitted, the other one is in a bound state of the dot.

The two-particle Hamiltonian reads:
\begin{eqnarray}
\label{hamiltonian}
& &{\bf H}(x_1,x_2)=-
\frac{\hbar^2}{2 m^*}\nabla^2_1 -  \frac{\hbar^2}{2 m^*}\nabla_2^2  \\ \nonumber
& &+ V_s(x_1) + V_s(x_2) +  \frac{e^2}{4\pi\epsilon \sqrt{\left(x_1-x_2\right)^2+d^2}}
\ ,
\end{eqnarray}
where $m^*$ and $\epsilon$ are the electron effective mass and the
dielectric constant of the material, respectively.   In particular  the calculations presented in this paper
have been performed using GaAs material parameters. The Coulomb term
includes the thickness a cut-off term $d$ that can be assumed to correspond roughly to the lateral dimension of the confinement \cite{Fogler}.
The spin degree of freedom does not enter into our calculation since
we neglect spin-orbit coupling
\footnote{ In fact, we assume that the two particles are
identical. This is justified if the two electrons are considered
spinless or if they have the same spin.  The only possible alternative, i.e. the two electrons having  opposite spin, corresponds  to
the simpler case of two distinguishable particles, not covered in the
present work. }.
As a consequence of the fermionic nature of the system we  impose
antisymmetrization  constraints under particle exchange to the  wavefunction
$\psi(x_1,x_2)$.  This is done by adopting antisymmetric boundary
conditions as briefly described in the following and detailed in
Ref.~\onlinecite{Berton}. 

The two-particle scattering state is  obtained by solving the
time-independent open-boundary Schr\"odinger equation ${\bf H}\psi = E
\psi$ in the 2D domain of interest.  To this aim we have used a numerical approach, based on a  generalization of the widely used 
\textit{quantum transmitting boundary method} \cite{Lent}. It  allows us to include
proper open boundary conditions and simulate the scattering of one electron
by a charge confined in a QD \cite{Berton}. In particular here we need four boundary conditions, one for
each edge of the square domain.  Since we have to impose exchange
symmetry to $\psi(x_1,x_2)$, they are equal in couples, apart from the
sign. In fact the form of the wavefunction when particle $1$ is in the
left lead ($x_1<0$) is
\begin{eqnarray} \label{boundaryleft}
& &\psi(x_1,x_2)|_{(x_1<0)}=
\chi_0(x_2) e^{ik_0x_1} +\\ \nonumber
& &+\sum_{n=0}^{M} b_n \chi_n(x_2) e^{-ik_nx_1} 
 \sum_{n={M}+1}^\infty b_n \chi_n(x_2) e^{k_nx_1}  ,
\end{eqnarray}
and, when it is particle $2$ the one in the left lead, the boundary
condition is $\psi(x_1,x_2)|_{(x_2<0)}=-\psi(x_2,x_1)|_{(x_2<0)}$.
For the other two boundaries, namely the conditions ($x_1>\Delta$) and
($x_2>\Delta$) it holds
\begin{eqnarray} \label{boundaryright}
& &\psi(x_1,x_2)|_{(x_1>\Delta)}= \nonumber \\
& &\sum_{n=0}^{M} c_n \chi_n(x_2) e^{ik_nx_1}
+ \sum_{n={M}+1}^\infty c_n \chi_n(x_2) e^{-k_nx_1} 
\end{eqnarray}
and $\psi(x_1,x_2)|_{(x_2>\Delta)}=-\psi(x_2,x_1)|_{(x_2>\Delta)}$,
respectively.

Let us now describe the above expressions.  We first define $T_n = E_{IN}
+ E_0 - E_n$ as the energy of an electron freely propagating in the
lead when the other one is bound in $\chi_n$ (a consequence of energy
conservation), and   $k_n=\sqrt{2 m^* |T_n|}/\hbar$. The right hand side  of Eq.~(\ref{boundaryleft}) (for the left boundary)  is the
sum of three terms.  The first one represents an electron  incoming as
a plane wave with energy $E_{IN}=\hbar^2 k_0^2 / 2m^*$, while the other
electron is in the QD ground state $\chi_0$.  The second term accounts
for all the energy-allowed possibilities with one electron bound in
the $\chi_n$ state,  and the other one reflected in the left lead,  with
wave vector $k_n$.  $M$ is the number of states for which $T_n$ is
positive. The third term accounts for the cases $T_n<0$, representing
the electron in the lead as an evanescent wave.  The right hand side  of
Eq.~(\ref{boundaryright}) (for the right boundary) has only two terms,
since the probability amplitude of a carrier incoming from the right lead is zero
in our system. The first term represents the $M$ energy-allowed
possibilities of an electron transmitted and freely propagating in the
right lead and the second term includes the outgoing-particle evanescent waves,
as in Eq.~(\ref{boundaryleft}).  The reflection and transmission
amplitudes in the various energy levels, $b_n$ and $c_n$ respectively, are unknowns and are
obtained by solving the Schr\"odinger equation with ${\bf H}(x_1,x_2)$
given by Eq.~(\ref{hamiltonian}) and with the two-particle energy
$E=E_{IN}+E_0$, imposing the two boundary conditions of
Eqs.~(\ref{boundaryleft}) and (\ref{boundaryright}).

In order to describe   the mechanisms characterizing   the  charge transport through a QD, 
and the kind of resonances showing up in transmission spectra, we  report in Fig.~\ref{fig:1}   the modulus and phase of the TC      as a function
of the initial energy of the incoming electron, for a given configuration  of the potential sketched in the  panel (a). 
In particular we consider a width $L$ of the potential well of 80 nm
with a depth $h$ of 150 meV. In this case  we have a single-channel scattering, i.e. the scattered particle  can have  a single energy,  while the bound  one is always
left in the ground state of the QD  well. When $E_{IN}$ is around 1.5 meV, the TC shows a  Breit-Wigner resonance, as shown in panel (b)\cite{Breit}. As it is well known, such  kinds  of resonances  stem
 from the coupling of a quasi-bound state to the scattering states in the leads and presents a Lorentzian line-shape whose amplitude $c_0$ 
is described by the  expression $c_0=C(i\Gamma/2)/(E_{IN}-E_{qb}+i\Gamma/2)$, where $C$ is a complex constant,
$\Gamma$ the width of the resonance, inversely proportional to the lifetime of the quasi-bound state with energy $E_{qb}$.
In particular, as we can see from   Fig.~\ref{fig:1}(b),  the  TC modulus (solid line) goes to 1 with a   symmetric Lorentzian peak
around the energy resonance, while the transmission phase (dashed line) smoothly changes by $\pi$.  
In addition  to the above resonance  two extremely sharp 
resonances for  $E_{IN}$   around 0.3  meV and 0.6 meV, are present.
 They are the so-called  Fano resonances (with asymmetric line-shape), which can  be ascribed to an effect brought  about by electron-electron correlation \cite{Fano}.
In fact they  originated  from the interference between  a resonant mechanism, due to the Coulomb blockade in the QD, and
a non-resonant mechanism, given by the transmission in the dot via a direct trajectory \cite{Fuh,Berton}.
In order to give a better insight into their properties we have reported in the top panel (c)  a zoom
of the   modulus and phase of the TC at an  initial energy of the incoming electron $E_{IN}$  around 0.3 meV. Here
the modulus    reaches  1 and  then goes to  zero, showing  the typical asymmetric Fano lineshape. 
The transmission phase  increases smoothly by $\pi$ on the resonance peak,  but shows an abrupt drop of  $\pi$ 
in correspondence of the zero of the transmission probability \cite{bertoni2}. We stress that the curves in  Fig.~\ref{fig:1}
correspond to a single-channel two-particle scattering, with the energy of the transmitted electron fixed by the boundary conditions.
As a consequence, no entanglement is generated between the two electrons here. The more interesting cases of two- and multi-channel
scattering will be presented in the following section.

In the last part of this section we describe the procedure we adopt  to evaluate the entanglement. In fact as  a consequence of the scattering,  quantum correlations  appear   
between the energy levels $E_n$ of the electron   bound in the potential well and the energies $T_n$ of the 
scattered  electron   allowed by  energy conservation.  Such an entanglement  may  be evaluated
from the transmitted  component of the two-particle wavefunction  
 in the right lead,    given by Eq.~(\ref{boundaryright}).
Obviously only the travelling components, with $n\leq M$ in the right hand side of Eq.~(\ref{boundaryright}),  must be considered,
since they are the only ones giving a non vanishing contribution to the current and also the only ones that could be revealed by measuring
 the energy of the electron propagating in the right lead. On the other hand  the evanescent
component of the wavefunction  plays  no direct role in the appearance of detectable 
 quantum correlations. In this case,  a  good entanglement measure  
is given by the von Neumann entropy of the reduced   density matrix $\rho_{Red}$,
obtained  by tracing  the two-particle 
density matrix $\rho=| \Psi^{tr}\rangle \langle \Psi^{tr}| /\langle \Psi^{tr}|\Psi^{tr}\rangle $ over the degrees of freedom of one of the  particles \cite{Peres},
where $\psi(x_1,x_2)^{tr} =
\sum_{n=0}^{M} c_n \chi_n(x_2) e^{ik_nx_1} $.

$\rho_{Red}$  will be  a $(M+1)\times (M+1)$ diagonal matrix  defined as 
\begin{equation} 
\rho_{Red}=\textrm{diag}\bigg[|\widetilde{c}_0|^2, \ldots, |\widetilde{c}_n|^2, \ldots, |\widetilde{c}_M|^2\bigg]
\end{equation}
with 
\begin{equation} 
 \widetilde{c}_n=\frac{c_n}{ \sqrt{\sum_{n=0}^{M} |c_n|^2}} .
\end{equation}
Thus the entanglement  can be expressed by means of the von Neumann entropy as
\begin{equation} \label{formu}
\varepsilon =-\textrm{Tr}\left[\rho_{Red}\ln{\rho_{Red}}\right]= -\sum_{n=0}^{M}  |\widetilde{c}_n|^2 \ln{|\widetilde{c}_n|^2}.
\end{equation}
$\varepsilon$ is bound in the interval $\left[0,\ln{\left(M+1\right)}\right]$ with $\varepsilon=0$  indicating no entanglement
and   $\varepsilon=\ln{\left(M+1\right)}$ indicating a maximally-entangled state.
We stress  that,  unlike other  works estimating the quantum correlation
in QDs \cite{Oli,Lopez},  we are not considering the transmitted and reflected component of the scattering wavefunction
as two different states that can be entangled with the QD. We estimate the entanglement between the dot and the \textit{transmitted} electron.
 In fact,  in our approach  the measure of the entanglement created in the system, 
evaluated by  means of Eq.~(\ref{formu}),  is not explicitely given in terms  of  the device transmission
and  reflection coefficients,  but it is a function of  the amplitude  probabilities of finding
the transmitted particle in one of the  possible energy states.
 Furthermore,  it is worth noting  that, although we do not use explicitely
the criteria developed to treat  the entanglement  of identical  particles \cite{Sch,bus,bus2},  the electrons in our system
are fully indistinguishable.  Nevertheless, since  the energy  of the incoming electron is not sufficient
to ionize the QD,  we may safely assume that   the scattered particle is  far enough from  the  one  left  in a bound state of the dot
 so that   the  overlap between their  spatial probabilities   is negligible. In other words we use the spatial position of the electrons
to  ``distinguish'' them, while the entanglement is between their energy states, as suggested in Ref. \onlinecite{eck}.

\section{Numerical results}\label{numerical}
In this section we  analyze the entanglement formation
in  a  two- or multi-channel scattering,   and its relation to the resonances  in  the TCs.
In particular,  it is  of interest to
study how the entanglement  depends upon  the kind of resonances,  since the detection of the latter, usually feasible through a current-voltage
 characteristic  of the quantum device,
can give information about the former.
In our approach we solve numerically 
the system for different  potentials  $V(x)$ (obtained by varying
the depth $h$ and the width $L$ of the  well), and  for different  initial energies  of the incoming electron $E_{IN}$.

\begin{figure}[h]
  \begin{center}
    \includegraphics[width=0.8\linewidth]{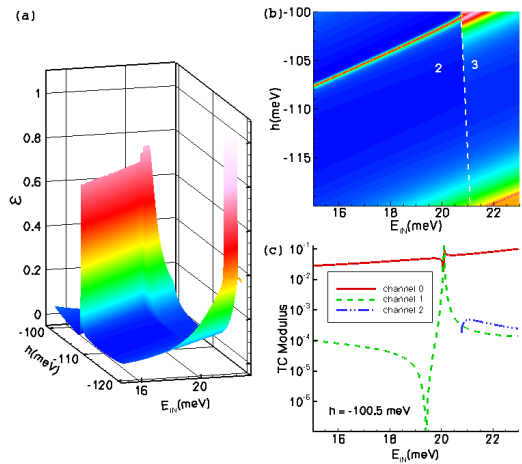}
   \caption{\label{fig:2} (Color online) Left panel (a):    entanglement  $\varepsilon$ as a function of the initial energy of the incoming 
electron $E_{IN}$ and of the depth of the  potential well of   40 nm.  Right panels: (b)    2D contour plot of the same data;
 the dashed line is a guide for the eyes and separates  the zone corresponding to two-channel scattering (left) from the one corresponding to a  three-channel scattering (right); (c) modulus of TC of the three channels
as a function of $E_{IN}$ with $h$ fixed at 100.5 meV. Note that the channel 2 is activated only at $E_{IN}\simeq 21 $ meV.}
 \end{center}
\end{figure}

In   panel (a) of Fig.~\ref{fig:2} we report
the entanglement  $\varepsilon$ of the system as a function of the initial energy of the incoming electron and of the depth
of the potential well,  whose width is kept constant at  $L$= 40 nm. The entanglement  presents  extremely sharp peaks  that, as  we shall see, 
 correspond to  the resonances  exhibited by the TCs of the various channels. From the 2D representation  of the same data, displayed  in Fig.~\ref{fig:2}(b), 
we observe that  the maxima of the entanglement spread in the region
corresponding to a three-channel scattering, that is separated by the dashed line from the one where the number of active channels is two. 
This suggests  us that also the kind of process, i.e. two- or multi-channel scattering, plays an important role into the entanglement formation.
The two cases  will be better analyzed  in the following subsections.

\subsection{Two-Channel scattering}
Here we study the creation of the entanglement when the incident particle, as a consequence
of the  scattering with the particle bound  in the QD, is transmitted with two possible energies, $T_0$ and $T_1$,  and, correspondingly, the final
QD energy can be $E_0$ and $E_1$.

\begin{figure}[h]
  \begin{center}
    \includegraphics[width=0.8\linewidth]{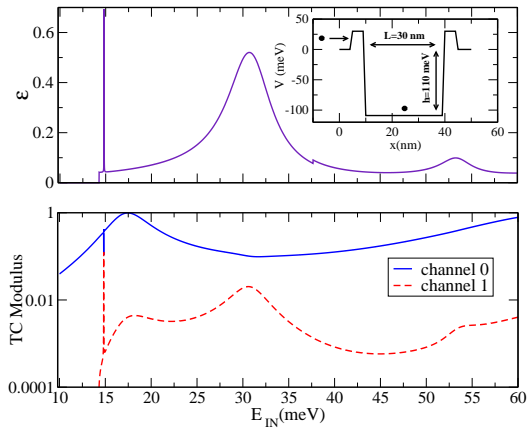}
   \caption{\label{fig:3} (Color online) Top panel: the entanglement  $\varepsilon$ as a function of the initial energy of the incoming 
electron $E_{IN}$. The inset displays the  potential $V(x)$  inside the scattering region: we consider
a potential well 110 meV deep and  30 nm wide. Bottom panel: the modulus of the TC of the  channels as a function of the initial kinetic energy of the incident electron:
0-th channel (solid line) and 1-st channel (dashed line). The scale of the abscissa is the same in the two panels.}
 \end{center}
\end{figure}

In the top panel of Fig.~\ref{fig:3} we report the entanglement as a function of  the  kinetic energy of the incoming particle,
for  the potential  sketched  in the inset of the figure. In particular 
we consider a width of the potential well  of 30 nm and  a depth of  $110$ meV. At  low energies the scattering
does not lead to the appearance of quantum correlations  between the two particles,  since only a single channel
is possible for the   transmission. It is therefore  possible to attribute
a specific  energy  to each particle: $T_0$ for the scattered electron and $E_0$ for the  bound one.  When $E_{IN}$   reaches a threshold value of about   14  meV
a new channel comes into play  as it can be seen  from the bottom panel  of  Fig.~\ref{fig:3}. There,  the dependence of the modulus
of the TC is reported  as a function of the initial energy of the incoming electron.  In correspondence to  
 the  energy at which the second channel is activated,   the entanglement shows a sharp increase.

\begin{figure}[h]
  \begin{center}
    \includegraphics[width=0.8\linewidth]{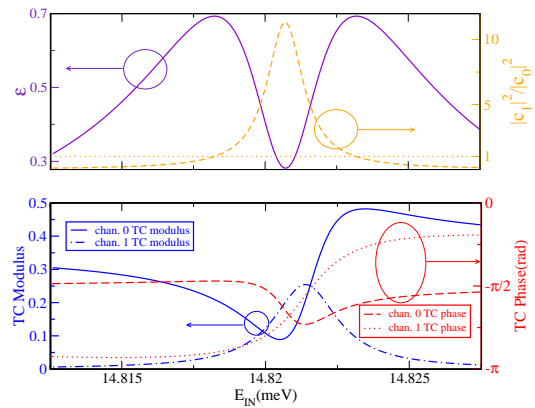}
   \caption{\label{fig:4}(Color online) Top panel:  the entanglement  $\varepsilon$ (solid line)
and   ratio between   the probabilities   of finding the scattered particle  in the  transmission channel 1, $|c_1|^2$,  and  0, $|c_0|^2$    (dashed line),
 against    the initial energy of the incoming 
electron   $E_{IN}$,  close
to a resonant condition. Note that, while $\varepsilon$ can range from 0 (no entanglement) to $\ln{2}$ (maximally entangled state), the bottom of the left
scale is set to 0.25 to  optimize the magnification effect in the figure. The horizontal dotted line is a guide for the eyes indicating 
the point  of equal scattering probabilities for the two channels, i.e. 1 in the left ordinates axis.
Bottom panel: TC of the two channels as a function of $E_{IN}$ around a Fano resonance. The solid and   the dash-dotted lines represent    
the modulus of TC of the  channel  0 and 1 respectively,  and refer to the left ordinate axis; 
the dashed and the dotted lines represent  the phase of the TC of  channel  0 and 1, respectively,  and  refer to the right axis.
 }
 \end{center}
\end{figure}

By comparing the two panels in Fig.~\ref{fig:3}, it is clear that  the resonances  of the TCs  play a key role in the entanglement formation.
In particular,   the behavior shown by the entanglement in correspondence of  a   Fano resonance is very different from
the one exhibited at  a Breit-Wigner resonance. When $E_{IN}$ is around 15  meV,  a
Fano resonance of the TC   is observed for channel 0. To better understand this phenomenon,
we report    in the top panel   of Fig.~\ref{fig:4} a  zoom of the curves of    modulus and phase (the latter was not shown in   Fig.~\ref{fig:3},  for clarity)
 of the TCs  of the two channels.
It is worth noting that, unlike the case of the  single-channel scattering, 
here the modulus of the TC attains small but non zero  values before the maximum, which, in turn,  results to be significantly lower  than 1. The TC of   channel 1
shows,  in correspondence of the Fano resonance of    channel 0, a Breit-Wigner resonance characterized by a phase change of about  $\pi$. In the energy
interval around 15  meV the  behavior of the quantum correlations, appearing in the systems as a consequence of the scattering,  is peculiar (see the  top  panel of  Fig.~\ref{fig:4}).
In fact when the  modulus  of the TC of the  channel 0 reaches its lowest value, the entanglement curve presents  a minimum. Such a minimum  is placed between two very  close maxima, 
where the entanglement (evaluated, as usual, by means of the von Neumann entropy
of the reduced density matrix)  is equal to $\ln{2}$. This  value indicates the condition of maximal entanglement in a two-channel scattering.
Such a condition is reached when  $|c_0|^2$ and  $|c_1|^2$, i.e. the probabilities that the scattering  occurs through the channel 0 or 1, respectively,  are equal, 
 and it implies that  the lack of knowledge about the state in the one-particle subspace is  maximum. We also report, in the top panel  of the Fig.~\ref{fig:4} (dashed line),
the ratio of the two transmission probabilities: the entanglement is maximum when the scattering probabilities in the two channels are the same, as indicated by the horizontal
dotted line drawn as a guide for the eyes.
What we found here  is in agreement with  previous analyses on  the two-electron  entanglement production
in two-electron systems for a two-channel scattering model, where both particles are injected in only one of two leads\cite{Oli,Lopez}. In fact, also in those cases,
the entanglement shows  a maximum when the transmission probabilities for the two  channels  are identical,
while it vanishes in correspondence of   the  single-particle resonances, where there is  no  uncertainty about the energy  of the particle.

 Figure~\ref{fig:3} shows  that the TC of the  channel 0 presents a  Breit-Wigner resonance for $E_{IN}\simeq 17.5$ meV.  Even if the modulus of the  TC  becomes equal to 1, 
the entanglement curve does not display  maxima or minima. This is due to the fact
that the Breit-Wigner resonance of the TC of    channel 0 does not influence the TC of    channel 1,
which does  not  show, at the specific energy,  resonances of any kind.  Therefore,   the scattering  phenomena taking place in the energy  interval around   17.5 meV 
do not play a special  role into entanglement formation of the two-particle system. Such  a behavior is in agreement with the one observed in 
other works,  showing that the maximal value of the conductance
does not always correspond to  the maximal entanglement \cite{Ryc}.

\subsection{Multi-channel scattering}
Let us  consider  now,  the  case of a multi-channel scattering (the scattered particle can leave the QD with more than two  energies),  and let us  investigate the  
different roles 
played by the Fano and  Breit-Wigner resonances in the entanglement formation.

\begin{figure}[h]
  \begin{center}
    \includegraphics[width=0.8\linewidth]{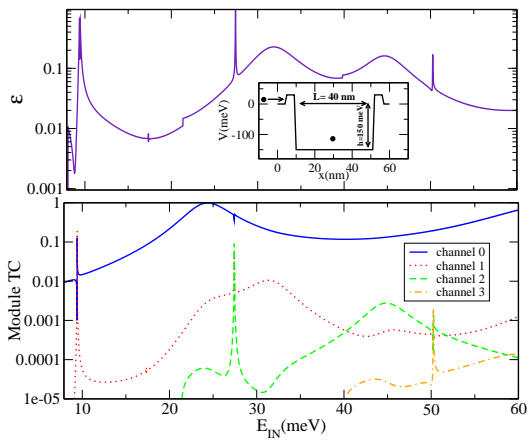}
   \caption{\label{fig:5} (Color online)  Top panel:   the entanglement  $\varepsilon$ as a function of the initial kinetic energy of the incoming 
electron $E_{IN}$. The inset shows  the profile of the potential $V(x)$ in the   scattering region: the  potential well is  150 meV deep and 40 nm wide.
Bottom panel: modulus of the TC of the  four channels as a function of $E_{IN}$:
channel 0 (solid line), 1  (dotted line), 2 (dashed line) and 3 (dash-dotted line). }
 \end{center}
\end{figure}

Here,  we consider   a  potential well   150 meV deep and  40 nm wide,
as reported in   the inset of Fig.~\ref{fig:5}.  For   $E_{IN}$ around 21 meV, the scattering passes from a two-channel to a three-channel process, 
and this transition is characterized
by a sharp increase of the entanglement. The same behavior    is found at    $E_{IN}=40$ meV,  where  an additional  transmission channel
 is switched on  (see Fig.~\ref{fig:5}). This is in agreement with the results of  the previous section and may be
considered  representative of a more general behavior, occurring whenever  a new  channel  becomes effective in the  scattering process.

\begin{figure}[t]
  \begin{center}
    \includegraphics[width=0.8 \linewidth]{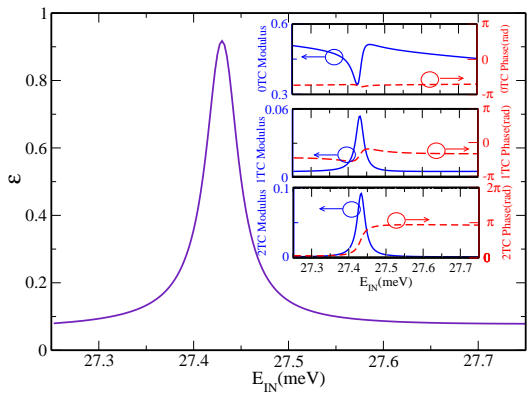}
    \caption{\label{fig:6} (Color online) The entanglement  $\varepsilon$ as a function of
      $E_{IN}$ around a Fano resonance of the TC of  channel 0in
      a three-channel scattering. The insets display the modulus (solid
      line, left axis) and phase (dashed line, right axis) of TC of three channels (0, 1, 2 from top to bottom) in the same energy range of the main graph.}
 \end{center}
\end{figure}

The quantum correlations appearing in the system, as a consequence of a multi-channel scattering,  
for  the energy values  around a Fano resonance show some differences with respect  to the  two-channel process.
In Fig.~\ref{fig:6} we report the curves of the  entanglement and, in the insets, the modulus and phase  of the   TCs of the three channels (0, 1, 2 from top to bottom) 
against   $E_{IN}$,  in the energy region  around  $E_{IN}=27.5$ meV,  where a Fano resonance  occurs for  channel 0. 
In fact, we observe, from the uppermost inset   that also in this case the  modulus  of the  TC has  a local minimum before reaching the  maximum.
The type of resonances of channels 1 and 2 (middle and bottom insets) cannot be clearly identified by
the module of their transmission coefficients, whose peaks are almost symmetric. However, it is clear for the transmission
phases, that channel 1 exhibits a Fano resonance, with the phase essentially unchanged through the peak, while channel 2 shows a Breit-Wigner
resonance, with a global phase variation of $\pi$. We stress that, unlike the two-channel scattering case, here the entanglement  does not show  a  minimum. 
Such a behavior can be ascribed to the fact
that, when the modulus  of the TC of    channel 0 is small,
the   TCs of the other two channels attain  values comparable to each other.
This  means  that  the probabilities of finding the scattered particles in those channels
are  almost   equal,  and  there is still a   lack of  knowledge about the state of the one-particle subsystem. Furthermore, 
we note that  here   the entanglement  presents a single  maximum whose  value exceeds  $\ln{2}$. Actually, the fact that   the number of degrees
of freedom is larger  than 2 increases the uncertainty  about the constituents of the system;  in fact Eq.~(\ref{formu}) gives a maximum value for the amount
of quantum correlations  that is larger for a larger number of possible states. For example, a system of two qutrits is able to attain
a larger value of $\varepsilon$ than a two-qubit system.
 \footnote{A qutrit is a unity of 
quantum information represented by a three-state system, in analogy with the qubit, embodied by a two-state fermion.}
The ability to tailor not only the degree of the entanglement, but also the number of possible states of the two
subsystems, by independently tuning $E_{IN}$ and the QD confining potential, could also  have implications beyond the theoretical estimation of the entanglement. The behavior described above
is repeated at  $E_{IN}\simeq 50 $ meV  where two Fano resonances
occur for the second and third channel in a four-channel scattering (Fig.~\ref{fig:5}).
It is worth noting that, in correspondence of a Breit-Wigner resonance in  the  channel 0 for $E_{IN}=25$ meV,
no additional  resonance  occur in the other channels and the entanglement does not present maxima or minima as it can be  clearly seen
from the upper panel of Fig.~\ref{fig:5}. Therefore  also in the case of multi-channel processes,
the Breit Wigner resonance   seems not 
 to induce sharp variations of $\varepsilon$.

\section{Conclusions}\label{Conclusions}
The   controlled production and detection of entangled particles in the solid state environment  represents
an experimental challenge. In this spirit,  various proposals for producing bipartite entangled fermionic systems  have been advanced,
on the basis of different physical mechanisms requiring a  direct interaction between particles \cite{bus,Oli,raa,bor2,Rasm,Lopez}. In this paper we have  investigated
the  quantum correlations appearing,  as a consequence of  a Coulomb scattering, between  two electrons having the same spin,  in a  system of physical interest, where the degree of the entanglement results to  be controllable  by  a proper tuning of the carrier energy and of the QD potential.  
Such a system consists of a  quasi 1D double-barrier resonant tunnelling device, where an electron
incoming from one lead  is scattered by the potential structure and, via the  Coulomb interaction, by  another electron bound in the QD.

The numerical procedure,  used to solve the model, is a generalization of the quantum transmitting boundary method \cite{Berton,Lent}.
It permits  us to obtain  the reflection and transmission  amplitudes of each scattering channel, 
for various configurations  of the potential, as a function of the initial energy of
the incoming electron. However,  we stress that, unlike the approaches followed by Lopez \textit{et al.}~\cite{Lopez} and by Oliver \textit{et al.}~\cite{Oli},
here we did  not use the reflected component of the scattered electron wavefunction to evaluate the entanglement of the two-particle system, but we estimated  the quantum correlations showing up
 between  the QD eigenstates and the  transmitted parts of the electron wavefunction. Furthermore,  this procedure   makes  it possible to investigate
the role played by the resonances of the transmission spectra  into the entanglement.
Although   our numerical analysis has been performed by using the GaAs material parameters,
they can be considered  representative of a more general behavior.

Our simulations  show  that the entanglement depends upon the kind of the resonance appearing in the  transmission spectrum.
A  single Breit-Wigner resonance is found  not to induce peculiar effects on  quantum correlations.
On the contrary,  in correspondence of  a Fano resonance of one of the TCs, not only  the other channels  TCs exhibit local maxima, 
but also the entanglement presents sharp peaks. Such a behavior can be related to the nature of the  resonances themselves.  In fact a Breit-Wigner
resonance is essentially a one-particle effect,   showing up also in a single-electron scattering. On the other hand,  a  Fano resonance  is  a genuine multi-particle phenomenon,
due to  electron-electron correlation,  and is not present for single-particle systems. 

Furthermore,  we showed that the appearance of
quantum correlations  in our system is also affected by the number of the  transmission channels, i.e. the number of possible energy levels
of the scattered particle (and, due to the energy conservation,  of the bound particle). In fact,  the entanglement  shows   a sharp increase
whenever a new channel is turned on. Moreover,  its behavior     for   energy values around  a Fano resonance is found to  depend
upon  the kind of process: two- or multi-channel scattering. For   the  two-channel case, 
the entanglement  presents  a minimum between two close maxima, which  indicate  the maximal uncertainty  about   the state  of the system. In the 
 multi-channel case,  a single maximum of the entanglement, with no minima, is observed.  When the energy levels of the scattered and bound  electrons  are only two,
the minimum of the entanglement is found in correspondence of the local minimum of the TC of the Fano resonant channel. In this case it maximizes   the 
possibility to ascribe specific energy states to the subsystems.
On the other hand,  for a multi-channel scattering,  a minimum of TC of a Fano resonant channel does not  imply a decrease  of uncertainty  about the subsystems, since the  TCs of the other  non Fano resonant channels
attain  values comparable to each other.

Finally,  the results of our paper suggest  that the manipulation 
 of Fano resonances and of the number of scattering channels may allow to significatively  influence   the degree  of  entanglement 
between  the transmitted electrons and the QD. A  promising  development of the present work could be the study of the entanglement in the case of  scattering of a single  electron by a few charges 
confined in the QD,  in connection with   experimental results obtained for  the coherent components of the transmitted current  in the case of  the multi-occupancy of the dot \cite{Kon,Aik}. In the latter
case, with three or more particles in the system, the spin degrees  of freedom cannot be factorized and their  inclusion in our approach, although quite feasible, results very challenging form the computational point
of view.

\end{document}